\documentclass[runningheads,a4paper]{llncs}
\usepackage{fancyheadings}
\usepackage{named}

\usepackage{amssymb}
\usepackage{graphicx, subfigure, multirow} 
\usepackage{amsmath}
\usepackage{mathtools}
\usepackage{natbib}

\usepackage{url}
\urldef{\maila}\path|wcasey@cert.org|
\urldef{\mailb}\path|shelmire@counterthreatunit.com|

\def\algo#1{ \small #1  \normalsize }
\newtheorem{lem}{Lemma}

\begin{document}

\mainmatter  % start of an individual contribution

\title{Signature Limits: An Entire Map of Clone Features and their Discovery in Nearly Linear Time.}

\titlerunning{Signature Limits.}
\authorrunning{W. Casey, A. Shelmire}

\author{William Casey%
\and Aaron Shelmire}
\institute{Carnegie Mellon University, Software Engineering Institute,\\
Dell Secure Works \\
\maila\\
\mailb\\
}

%\institute{Carnegie Mellon University\maila, DELL Secure Works\mailb}

%\toctitle{Signature Limits: An Entire Map of Clone Features and their Discovery in Nearly Linear Time.}
%\tocauthor{Authors' Instructions}
\maketitle

\begin{abstract}
We address the problem of creating entire and complete maps of software code clones (copy features in data) in a corpus of binary artifacts of unknown provenance. We report on a practical methodology, which employs enhanced suffix data structures and partial orderings of clones to compute a compact representation of most interesting clones features in data. The enumeration of clone features is useful for malware triage and prioritization when human exploration, testing and verification is the most costly factor. We further show that the enhanced arrays may be used for discovery of provenance relations in data and we introduce two distinct Jaccard similarity coefficients to measure code similarity in binary artifacts. We illustrate the use of these tools on real malware data including a retro-diction experiment for measuring and enumerating evidence supporting common provenance in {\it Stuxnet} and {\it Duqu}.  The results indicate the practicality and efficacy of mapping completely the clone features in data. 
\end{abstract}

\section{introduction}
In 2011 the security community identified a relation of provenance between the {\it Stuxnet} and {\it Duqu} malware families \cite{Chien:2012}.  
The relation was substantiated by laborious reverse engineering digital artifacts\footnote{Artifacts are malware binaries, files, or digital evidence of a computer/network attack.} which reveled compelling evidence of code sharing.  
These reports addressed the underlying question of provenance in malware but left in question how much code sharing took place and further whether computational methods could be designed to measure and detect code sharing.  
%In 2010 several studies focused on the {\it Zeus/Zbot} malware samples with rigorous investigations of single examples \cite{Binsalleeh:10,Ormerod:2010,Nayyar:11}; while the analyses exposed the mechanics of these samples, the question of code associations between samples or even associations to the pro generating code remains open.  
%In addition the value of such associations remains unassessed.
%  We show that investigating associations leads to surprisingly simple understandings.

Scalable methods to triage and cluster malware using signatures have been considered in \cite{Bayer:2009},\cite{Brumley:BitShred},\cite{Kang:MAST} and \cite{Lakhotia:vilo}; however each of these methods employ lossy data reductions.  While these methods focus much attention on understanding error rates to achieve scalability, they leave open the question of whether the the tradeoff between statistical power and scalability is necessary to achieve clustering methodology.   

These problems provide a high level view of contemporary efforts in cyber security.  Common to both problems is the need to identify and map all common strings or shared code segments termed {\it code clones} within a limited set of artifacts or against a reference data set of known artifacts.  
Calling on recent advances in suffix-data structures and succinct data structures, we consider efficient computational methods for mapping {\it code clones} (all copy features in data) which are both complete for provenance studies and compact enough to scale to large clustering problems.

Tree and Array construction and merge these advances with a practical model for exact code clones leading to practical methods for malware identification and triage and prioritization of reverse engineering resources.  
\subsection{Background.}
This effort merges ideas from several distinct areas including: mathematics of measure theory, algorithm design calling on advances in suffix data structures, software engineering research which has recently suggested modes and models for code cloneage, and cyber security research which provides the motivating problems.

Code clones have been discussed in the area of software engineering where clones arise from a limited number of generating events including copy and paste, code reuse, common authorship, derived or augmented data, common linked artifacts, etc.
For large software projects code cloning is an important factor for software maintenance and while the engineering benefits of cloning are debated there is general agreement that identifying clones is an important capability \cite{Kim:2005}.
Clones can be efficiently identified in large-scale software projects \cite{Kamiya:2002} \cite{Li:2004} where commercialized  products have been developed. 
Recently modeling clone evolution has become an active area of research \cite{Antoniol:2002} \cite{Livieri:07}.  
Definitions of code clones vary across the literature and are a developing area of research (see \cite{Kim:2005} and \cite{Roy:07} for surveys). 
Our notion of code clones (presented in the next section) is novel and designed to both model gross structural features within the corpus using few quantities and be computable with suffix data structures.  We present a mathematical description of a measure space making our notion of code clone comparable to all other formal notions.

The question of how to organize and represent a text corpus for optimized retrieval and search has been motivated by diverse problems in areas of information retrieval \cite{Amir:1994}, \cite{Blumer:1987} and \cite{Ferragina:95}, pattern matching \cite{Weiner:73}, software analysis \cite{Baker:93}, and bio-informatics \cite{BieganskiRCR:94} \cite{Gusfield:1997} where there are several well developed techniques based on suffix trees \cite{McCreight:1976} \cite{Ukkonen:85}, compressed suffix trees \cite{Navarro:2007}, and suffix arrays \cite{Manber:1990} and \cite{FM:04}.  %\cite{Itoh:1999,Kim:2003,Ko:03,Manber:1990,FM:04,Siren:09}.  
%Alternatives to suffix tree/arrays have been developed using modified B+ trees \cite{Ferragina:95} to index text.
In addition indexing for dynamic data sets \cite{Amir:1994,Ferragina:95} has been reported.  Significant to very large data sets are \cite{Ferguson:Femto} where researchers have considered applicationions of suffix data structures to data at scale. %\cite{Bingmann:FO13,CrauserFerragina:99,Ferguson:Femto,Ferragina:95,Navarro:2007} 

In addition to the identification of longest common substrings (LCS), statistical analysis of the content space has been suggested \cite{apostolico:2003} but not developed to the extend that useful code-clones can be identified in malware artifacts.%\cite{Apostolico:85,Gusfield:1997,apostolico:2003}
%Our contribution is focused on increasing these statistical capabilities to the suffix-tree data structure extended to corpora (sets of files).  
While the topic of extending index-recallers to corpora is addressed in \cite{BieganskiRCR:94}, \cite{Blumer:1987}, \cite{Ferragina:95} and \cite{Gusfield:1997}, with emphasis on suffix tree being central in \cite{BieganskiRCR:94} and \cite{Gusfield:1997}, our contribution develops tree-traversal and indexing arrays for quantities of {\it entropy}, {\it length}, {\it multiplicity}, and {\it file coverage} needed for discovery (i.e. ``calling'') of clones in software executables.  

We further consider methods to represent a set of clones that is both compact and complete.
The use of suffix-trees for the analysis of set-algebra of corpus indices has been studied from a formal concept analysis and data-mining approach in \cite{Ferre:07} where suffix trees are implemented to identify {\it string-scales} specific to a set lattice.

We show that this merger of a mathematical measure space for code clones combined with enhanced and tailored suffix data provides effective applications to problems in cyber security addressing provenance studies and data clustering.
\section{Definitions and Clone Model.}
For a string $\lambda$ over a finite alphabet $\Sigma$, let $| \lambda | $ denote the string length, $\lambda[j] \in \Sigma$ the $j$th symbol,  and $\lambda[j:k]$ the substring $\lambda[j]  \lambda[j+1] \hdots \lambda[k-1]$.
A {\bf corpus} is an ordered set of strings $\Omega = \{ \omega_0 , \omega_1, \hdots , \omega_{n-1} \}$ over a common finite alphabet $\Sigma$, therefore each string $\omega_i \in \Sigma^*$ for $i \in \{ 0, 1, \hdots, n-1 \}$.  
The corpus size is measured by the number of strings $| \Omega | = n$, and the total length of corpus $|| \Omega || =  \sum_{k = 0}^{n-1} | \omega_k | $.
% average length of member $\langle \Omega \rangle = \frac{1}{|\Omega|} \sum_{k = 0}^{n-1} | \omega_k | $.

%A {\bf corpus offset} is any integer position pair $( i, j )$ meeting the condition that $i \leq |\Omega|$ and $ j \leq |\omega_i|$.  The set of all corpus offsets is represented by the set ${\cal O} = \{ (i,j) : i < |\Omega| ,  j < |\omega_i| \}$.  

A {\bf corpus region} is represented by a tuple $(i, j , k)$ with $i < | \Omega| $ and $0 \leq j \leq k < |\omega_i|$;  the first index specifies the corpus element  from which the region is drawn (i.e. a given string $\omega_i$), while the second and third indices provide the region within string $\omega_i$ beginning with and including offset $j$ and covering up to but not including offset $k$.  
Associated with each corpus region $(i, j, k)$ is the sub-string: $\omega_i[j: k] \in \Sigma^*$.
Let ${\cal R}$ denote the set of all corpus regions: ${\cal R}= \{ (i, j, k) \ |\  i < |\Omega|, 0 \leq j \leq k < |\omega_i| \}$.  
Assume the following functions: $\text{\sc{File}}((i,j,k)) = i$, $\text{\sc{Offset}}((i,j,k)) = j$, and $\text{\sc{End}}((i,j,k)) = k$ providing the coordinate projections for tuples in ${\cal R}$.

We identify the relation between corpus-regions and observed sub-strings by the {\bf content map}:
$$
\Gamma : {\cal R} \rightarrow \Sigma^* : \{ (i, j,k) \} \rightarrow \omega_i [j, k].
$$
We refer to the inverse of $\Gamma$ as the {\bf region recaller}; for any string $\lambda \in \Sigma^*$ a subset of matching corpus regions is returned: 
$$
\Gamma^{-1} : \Sigma^* \rightarrow 2^{\cal R}  : \lambda \rightarrow \lambda^{-1} ( \Omega ),
$$
with 
$$
\lambda^{-1} ( \Omega ) = \{ (i, j, j + | \lambda | ) \in {\cal R} : \omega_i[j: j+ |\lambda|] = \lambda \}.
$$
If a string (over $\Sigma$) is not observed in the corpus, the region recaller returns the empty set denoted $\emptyset$.  
The inverse of the empty string $\epsilon$ can be defined as ${\cal R}$ without loss of generality or specificity.  
With $\Omega$ (and consequently $\Gamma$) fixed, we refer to $\lambda^{-1}( \Omega )$ as the {\bf pullback} and denote it as $\lambda^{-1}$ for short.  The pullback of $\lambda$ returns the corpus regions where the string $\lambda$ is found.
%shorthand for $\lambda^{-1}(\Omega) = \Gamma^{-1} ( \lambda ) \subset {\cal R}$.   

The {\bf observed language of the corpus} is the set of all strings with non-empty pullback: 
$$
{\cal L}( \Omega ) = \{ \lambda \in \Sigma^* \ | \ \lambda^{-1} \not = \emptyset \}.
$$
%We also use the {\bf statistical indicator} to identify subsets of strings, for $A \subset \Sigma^*$,
%$$
%1_A : \Sigma^* \rightarrow \{0,1\} : \lambda \rightarrow \begin{cases} 1 \text{ if } \lambda \in A, \\ 0 \text { otherwise }.\end{cases}
%$$
\subsection{Mathematics of Clones.}
Clones are the content strings found in multiple locations of a corpus; they provide introspection and discovery opportunities for uncharacterized data.  In order to concretely discuss clone concepts we describe clones mathematically as set systems in ${\cal L}(\Omega)$.  
Let $\Omega$ be a fixed corpus; we will use $\lambda^{-1}$ to mean the pullback $\lambda^{-1} ( \Omega )$ for any $\lambda \in \Sigma^*$.  
We start by introducing simple notions of code-clones and discuss how the different notions relate as nested sets.  Next, we focus on statistical features of cloneage needed to be effective in malware discovery.  
Toward these goals we add additional qualifiers to enrich the concept of code-clones.
We present a general nested model of cloneage in four parameters that will be used in applications for malware clone mapping, discovery, and measures.
We indicate the underlying mathematics of this model and justify why we chose these clone quantities.

{\bf Simple Clone Concepts:}
A simple notion of {\bf code-clone} is any snippet of code identified in multiple locations or in multiple files.  Two definitions capturing these notions are: 
$$
\text{M-Clone} = \{ \lambda \in \Sigma^* : | \lambda^{-1} | > 1 \}, $$ 
and
$$
\text{F-Clone} = \{ \lambda \in \Sigma^* : | \{ \text{\sc{File}}( x ) : x \in \lambda^{-1} \} | > 1 \}.
$$  
Note the dependencies in these models as $( \lambda \in \text{F-Clone} )\Rightarrow (\lambda \in \text{M-Clone})$.  While the statement $| \{\text{\sc{File}}( x ) : x \in \lambda^{-1} \} | > 1$ is sufficient for $ | \lambda^{-1} | > 1 $, it is not necessary as $\lambda$ may be found in each file of the corpus but never found duplicated at multiple offsets within any file.

Both of these sets are efficiently accessible using Suffix Trees \cite{Gusfield:1997}; however, for the task of malware discovery these notions are ineffective because a large volume of {\sc M-Clone} and {\sc F-Clone} may include byte padded sequences.  Thus, additional considerations including the statistics of entropy are needed to distinguish a more interesting set of clones for discovery, triage and analysis.

To further generalize the notion of code clone we consider statistical measures of string content, such as the Shannon Entropy function and how it may qualify clones.  
Let $\lambda \in \Sigma^*$, for $v \in \Sigma$; let $X_v ( \lambda ) = | \{ j < |\lambda| : \lambda[j] = v\} |$ be an observed symbol count, and let $\theta_v (\lambda) = \frac{ X_v (\lambda) }{ | \lambda |  }$ be the normalized symbol frequency.  The {\bf Entropy} for $\lambda$ may be defined as $H(\lambda) = \sum_{ \theta_v > 0 } \theta_v \log { \frac{ 1 }{ \theta_v } }$.

Using the entropy function, we obtain a more useful set of clones by conjoining a lower entropy threshold to clone criteria, that is:
$$
\text{M-Clone}_h = \{ \lambda \in \Sigma^* : ( | \lambda^{-1} | > 1 ) \wedge ( H ( \lambda ) > h )\}, 
$$ 
with the associated multi-file clone class as:
$$
\text{F-Clone}_h = \{ \lambda \in \Sigma^* :  ( | \{ \text{\sc{File}}( r ) : r \in \lambda^{-1} \} | > 1 ) \wedge ( H( \lambda ) > h ) \}.
$$
This extension to the clone model provides selectability against low entropy strings such as null byte pads\footnote{Zero padding a section of data is a common technique for file formats.} which are common in binary artifacts; however, they are accidental clones which we must regard as uninteresting.  In our experiments low entropy strings are often the longest common substring (LCS) and therefore a parameter such as $h$ is necessary to recover meaningful signals from suffix-arrays.  

In addition to {\it clone entropy} $H(\lambda)$, we further extend the concept of clones to include quantities of {\it clone length} denoted as $D(\lambda) = | \lambda |$, {\it clone multiplicity} denoted as $C(\lambda) = | \lambda^{-1} |$, and {\it file coverage} denoted as $F(\lambda) = | \{ \text{\sc{File}}(r) : r \in \lambda^{-1} \}| $,

{\bf Clone Model:}
We arrive at a {\bf general model of clones} over the content ${\cal L}(\Omega)$ by letting the tuple $\langle d, h, f, c \rangle$ represent the following subset of ${\cal L}( \Omega ) $:
$$
\langle d, h, f, c \rangle = \{ \lambda \in {\cal L}( \Omega ) : ( D(\lambda ) > d ) \wedge ( H(\lambda ) > h ) \wedge ( F( \lambda ) > f  ) \wedge ( C( \lambda ) > c ) \}.
$$
Letting variables $d,h,f,c$ range freely we have described a {\bf clone class} within the context of the partial ordering of $2^{{\cal L}( \Omega )}$ by sub-set containment.

For $\langle d,h,f,c \rangle , \langle d',h',f',c' \rangle \in 2^{ {\cal L}(\Omega) }$, we have the following nesting property:
$$
\langle d',h',f',c' \rangle \subseteq \langle d,h,f,c \rangle \Leftrightarrow ( d' \geq d ) \wedge ( h' \geq h ) \wedge ( f' \geq f ) \wedge ( c' \geq c ).
$$
Using the clone class in quantities $ d, h, f,c$ we may organize our simple clone concepts with set inclusion indicated by arrows as follows:
$$
\begin{array}[c]{ccccc}
\text{M-Clone}_{h} &= \langle 0,h,1,0 \rangle &\stackrel{ }{\leftarrow}& \text{F-Clone}_h &=\langle 0,h,0,1 \rangle\\
&\downarrow&&&\downarrow\\
\text{M-Clone} &=\langle 0,0,1,0\rangle &\stackrel{ }{\leftarrow}&  \text{F-Clone} &=\langle 0,0,0,1 \rangle\\
&\downarrow&&&\\
{\cal L}(\Omega) &= \langle 0,0,0,0 \rangle &&&\\
\end{array}
$$
%We later revisit some extensions to the lattice diagram and indicate some parameters useful for finding code-clones including a method to reconcile with data ordering.
%Notice the dependence $c(\lambda) \geq f( \lambda)$, for example consider a content string $\lambda$ found in $k$ files (i.e. $F(\lambda) = k$) then we have $C(\lambda ) \geq k$ because $\lambda$ could occur at more than one offset in one or more of the $k$ distinct strings of the corpus (i.e. files)
The clone class organizes the collection of clone sets in ${\cal L}(\Omega)$ into a nested family of {\it cylinder sets}.
% which are used to develop probability measures on clone concepts (in parameters $d,h,f,c$ ).
Cylinder sets (with set subtraction) may construct sets with each quantity bounded below and above; for example $\langle d_1,h_1,f_1,c_1 \rangle \setminus \langle d_2,h_2,f_2,c_2 \rangle $ specifies $  \{ \lambda \in {\cal L}(\Omega) : ( d_1 \leq D(\lambda) < d_2 ) \wedge ( h_1 \leq H(\lambda) < h_2 )  \wedge ( f_1 \leq F(\lambda) < f_2 )  \wedge  ( c_1 \leq C(\lambda) < c_2 )   \}$.  
Two-sided bounds for each quantity provide a richer class of clones; we will see that two sided bounded quantities are also computable with a single pass over the suffix data structures in the {\sc Traverse-Tree} method presented in section 3.3.
Closure under set operations (union, intersection, complement) of the cylinder sets generates a sigma algebra and therefore provides a mathematical measure space.

\paragraph{Justification of Clone Model.}
Closure of our clone class $\{ \langle d, h, f, c \rangle : d \geq 0, h \geq0, f \geq 0, c \geq 0 \}$ with set negation, intersections (conjunctions), and unions (disjunctions) generates a sigma-algebra ${\cal C}$ which is a coarsening of $2^{{\cal L}(\Omega)}$.
Therefore our notion of clones provide a measure space: $\langle {\cal L}(\Omega), {\cal C}\rangle$ which approximates $\langle {\cal L}( \Omega ), 2^{{\cal L}(\Omega)} \rangle$ and may be compared to other formal notions of clones.
Although the dimension of this clone model is low with only four free variables we shall argue that these are simple to build into suffix array indices and sufficient for calling interesting sets of clones from malware artifacts. Further the low dimensionality reduces the search for features of a corpus quantified as regions in the parameter space of $d,h,f,c$.  
% we continue to focus the presentation on the efficient computing of members of ${\cal C}$, efficacy of ${\cal C}$ to capture important concepts of clones in malware data.  We provide a brief description of how we arrived at these quantities as a sparse-representation of code clones in a corpus.  We argue that the quantities of the clone model are expressive for clone discovery in uncharacterized data including large variations in string entropy.  
Functions {\it file coverage} $F$ and  {\it clone multiplicity} $C$ are monotonically non-increasing in the suffix-order relation on ${\cal L}(\Omega)$; that is to say, if $\zeta$ is a suffix of $\lambda$ then $F(\zeta) \geq F(\lambda)$ and likewise for $C$.  However as mentioned above they measure different notions of cloneage with ratios expressing a comparison of self-similarity to similarity in the corpus at large.
The {\it clone length} function $D$ is monotonically increasing in the suffix-order relation as $D(\zeta) < D(\lambda)$.  
Therefore setting minimum values of $d,f,c$ works to select clones from a corpus by using opposing criteria in the suffix-order on ${\cal L}(\Omega)$.
The {\it clone entropy} function has no monotonic property in the suffix-order but is effective in selecting against low string entropy.  
Entropy selection is useful for executable modules which display wide variations including common null byte sequences.

\section{Methodology:  clone calling with arrays and representation for clone sets.}
%Suffix Arrays supporting measures on cloneage and representation 
The main result of this section is that we adapt suffix trees/arrays to {\it call} or enumerate the members of $\langle d,h,f,c \rangle$ in time: $O( ||\Omega|| \log{( | \Omega | )} )$.  
%and more generally identify events in the measure space $\langle {\cal L}(\Omega), {\cal C} \rangle$ in
We further show that clone sets $\langle d,h,f,c \rangle$ are reducible to a much smaller subset called a {\it max-clone representation} by use of a suffix-relation on $\Sigma^*$.  
The max-clone representation admits to both meaningful visualizations and application of measures to identify and infer provenance in artifacts (discussed in the next section).
We present a brief historical development of suffix data structures, subword trees, and arrays to discuss the {\sc Traversal-Tree} procedure which produces the arrays enhanced with clone quantities.  
Our model for suffix data structures is Ukkonen's suffix tree \cite{Navarro:2007},\cite{Ukkonen:85} and \cite{Ukkonen:95} and we follow its terminology and developments; for further background we suggest \cite{Navarro:99}.  Since suffix arrays may emulate suffix trees \cite{Abouelhoda:2004} our method is possible for various suffix array implementations as well.  
To be as general as possible we describe the minimum data requirements of suffix-tree nodes to complete the {\sc Traversal-Tree} method.
%we also provide background on suffix arrays, tree emulation, succinct data structures, and stringomics.
\subsection{Suffix Data Structures.}
Given a set of strings $S$, an index tree ({\it trie}), such as the {\it PATRICIA trie} \cite{Morrison:PATRICIA}, is a tree graph which encodes a finite state automaton (FSA) for acceptance of any input matching a member of $S$.  Each string from $S$ corresponds to a path from the {\it root} node to a {\it leaf}, and paths are merged by shared prefixes to form a trie (tree index). As an FSA, this structure may be considered an Aho-Corasick string matcher.
%While there may be more space-efficient FSA to construct for that problem, the {\it trie} performs well with problems associated with redundant or large number of queries against a relatively static set of strings.  
%This data structure is often utilized as a Aho-Corasick string matcher or a {\it Patrica trie}; after construction the performance of search is limited only by length of input query.

The set of all suffixes of string $\omega$ is denoted by ${\sigma}(\omega)$ and defined as: ${\sigma}(\omega) = \{ \omega[0:k] : k \in \{ 0, 1, \hdots |\omega| \} \}$. 
A {\bf Suffix trie} for $\omega$ is constructed by creating a {\it PATRICIA trie} on $\sigma(\omega)$.  

%The Patricia Trie for $\sigma( \omega )$ is a {\bf suffix trie} for string $\omega$.

The {\it suffix trie} may be used as an entire index of all substrings because any substring of $\omega$ can be written as a prefix of a member of ${\sigma}( \omega )$.  
Further the tree structure is meaningful for the problem of content mapping as the internal branching nodes of the structure are in correspondence with redundant strings of the text, the deepest internal branch of which is called the {\bf Longest Common Substring} (LCS) \cite{Gusfield:1997}.
\begin{center}
\begin{figure}[h!]
  \begin{center}
        \subfigure[]{
          \fbox{ \includegraphics[angle=90,width=54mm]{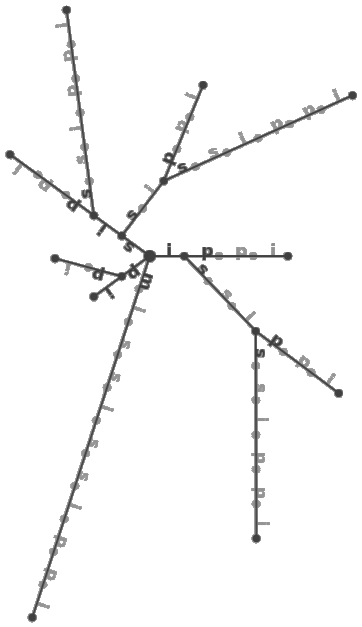} }
          \label{fig:A0a}
        }
        \subfigure[]{%
          \fbox{ \includegraphics[angle=90,width=54mm]{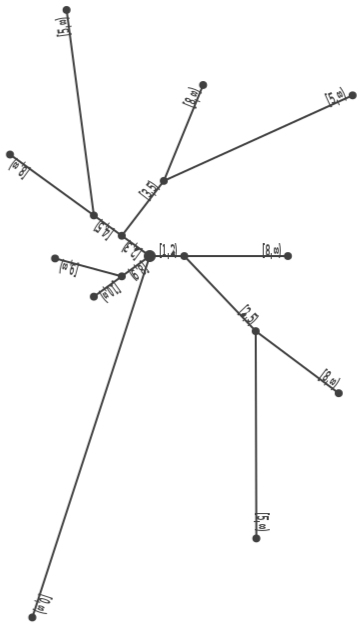} }
          \label{fig:A0}
        }
  \end{center}
  \caption{%
(a) Suffix trie and tree for $\omega =$$'mississippi'$; states of the suffix trie are indicated by nodes and state transitions by edges labeled with letters.  Ukkonen's suffix tree only requires explicit states (black) and is able to emulate the implicit states (gray) of the trie. %  is a PATRICIA trie for ${\sigma}(\omega)$ represented by either black or grey,  The black circles are explicit states while the gray circles are implicit states in Ukkonen's Suffix Tree . 
The tree {\it root} node is in the center and the set ${\sigma}(\omega)$ is displayed in lexicographical order starting at angle $\frac{\pi}{2}$ and rotating $2\pi$ in a clock-wise direction.  Notice also the deepest branching node in the tree corresponds to longest common string $'issi'$.
(b) Suffix tree for $\omega =$$'mississippi'$.  Replacement of the transition labels with offsets and length indices (referencing the input string $\omega$) create the suffix tree.  In addition each node maintains a set of children branches and a suffix pointer (not shown).
     }%
  \label{fig:A}
\end{figure}
\end{center}
The suffix trie data structure admits to a compact representation by removing internal non-branching nodes and emulating transition-labels for implicit states (see figure \ref{fig:A0a}).  
Further there is no need to store transition-labels as they can be recovered from offsets (in $\omega$), further reducing the space requirements for suffix tree nodes (see figure \ref{fig:A0}).

For fixed and finite alphabet $\Sigma$, the resulting data structure is linear in space $O(|\omega|)$ and constructed in linear time $O(|\omega|)$ \cite{Ukkonen:95}.  
%An additional feature of the construction in \cite{Ukkonen:95} is that the data structure is constructed online, meaning that at the $k$th step the index-recaller for $\omega[.k]$ is available.  
Further the data structure can be traversed in linear time $O(|\omega|)$ to identify the deepest branching node and equivalently the LCS of the text (see \cite{Gusfield:1997} for additional details).

%Following Ukkonen's procedure of removing non-branching nodes (\cite {Ukkonen:95}) we arrive at the suffix tree data structure emulating a PATRCIA trie for ${\cal S}(\omega)$.  
%The Suffix tree for ${\cal S}(\omega)$ is illustrated below as a compressed PATRCIA trie (with implicit states shaded) and with its more specialized and equivalent offset-interval augmentation of Ukkonen\cite{Ukkonen:95}.
In addition to trees, suffix arrays are constructed in near linear to linear time \cite{Karkkainen:2006} \cite{Kim:2003,Ko:03} and may emulate suffix trees \cite{Abouelhoda:2004}; therefore what can be performed on Ukkonen's tree extends in principle to many array implementations as well.
More recently, succinct data structures have achieved greater compression of suffix trees and arrays for lossless index re-callers \cite{FM:04} \cite{Navarro:2007};  for example the Ferragina Manzini index structure (FM Index \cite{FM:04}) utilizes the Burrows-Wheeler transform to compress the suffix array in memory.  
%In addition suffix arrays can emulate suffix trees (see \cite{Abouelhoda:2004}).

\subsection{Implementation: Construction of a Suffix Tree for Malware Artifacts.}
In order to scale the suffix tree beyond system external memory (EM) data structures are possible \cite{Arge:96} and \cite{Ferragina:95}.  
Beginning from Ukkonen's suffix tree algorithm, we implemented an external memory set of c-programs for a corpus over the bytes alphabet $\Sigma = \{ 0, \hdots, 255 \}$.  
%CQEST is also designed to study external memory management techniques for both build and traverse (next section).  
%Converting an offset in $\omega$ to a corpus region could either be done by including file offsets for leaf nodes (as we do) or by maintaining a corpus offset table \cite{Gusfield:1997}.
%We are able to comfortably build trees well beyond memory capacity, and our current efforts are focused on improving memory management and compression of data structures.  We are currently able to process the problems in our application section (including visualization) within minutes on a modest laptop computer.  
%%Our work with data structures is not yet complete so we postpone the discussion of performance and scale but suggest the many references we have previously cited in the context of external memory and very large data sets which indicate the adaptation of these data structures to large scale is feasible.
% and can contribute the clone model and analytics to the discussion of thwarting malware attacks.
% as all applications presented in the next section are performed on laptops within minutes.
%We augmented Ukkonen's suffix tree algorithm and were able to construct our suffix trees for data sets as large as 100Gb but consider the problem of optimizing memory management strategies for build and traverse somewhat difficult.  We also augment the suffix tree node data structure with a file index (at leaf nodes) for support of our corpus indexing.

We demonstrate an externalized version of Ukkonen's suffix tree algorithm augmented to support corpus indexing (by adding file index and local offsets within the file) to leaf nodes.  
Below in Figure \ref{fig:large_scale_data} we visualize a suffix tree data structure constructed to analyze string structure in {\it Aliser File Infector} malware artifacts.  
\begin{center}
\begin{figure}[ht!]
        \subfigure[zoom x10]{%
            \label{fig:first}
             \fbox{\includegraphics[angle=0,width=54mm]{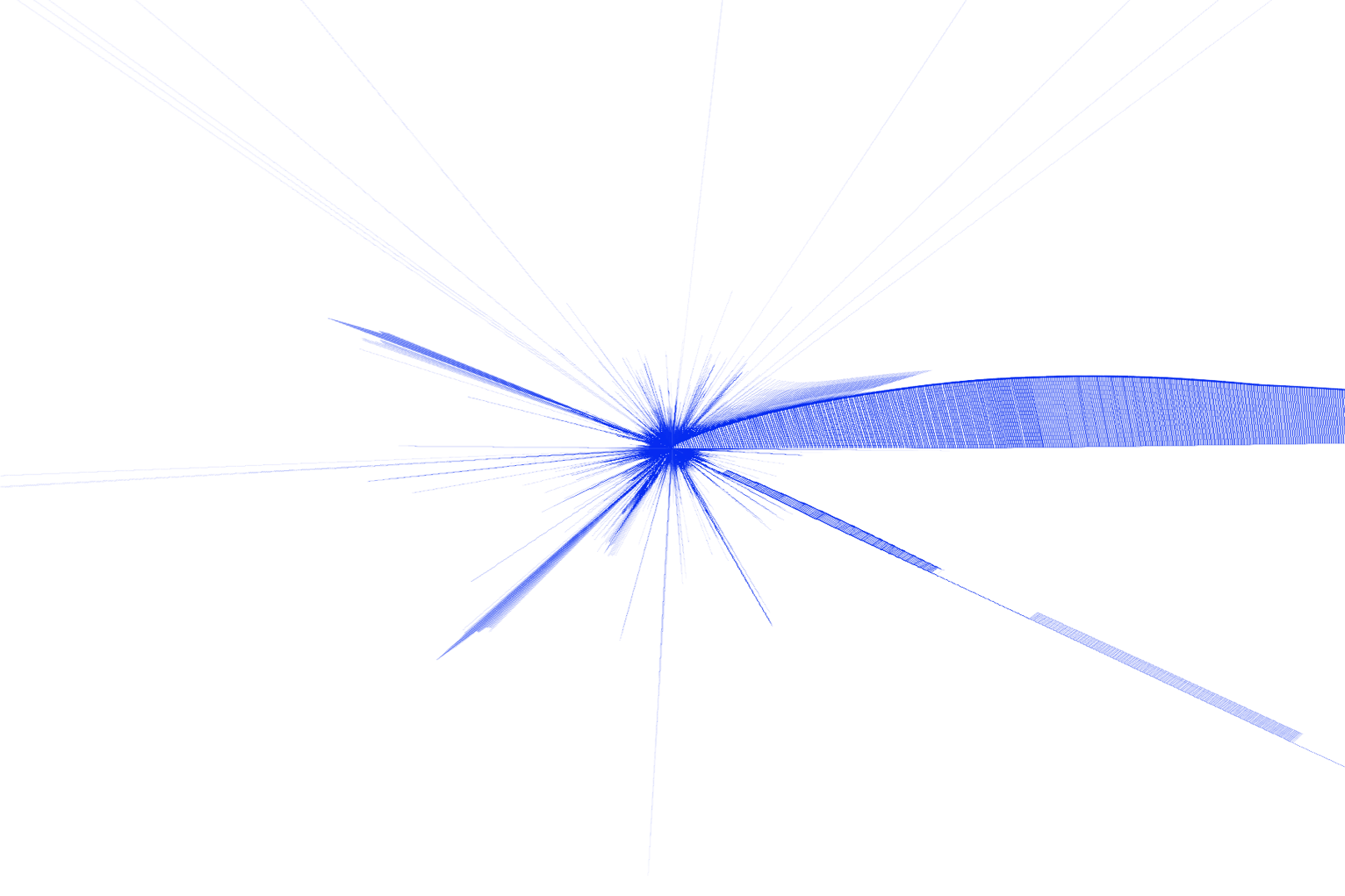}}
        }%
        \subfigure[zoom x100]{%
           \label{fig:x10}
            \fbox{\includegraphics[angle=0,width=54mm]{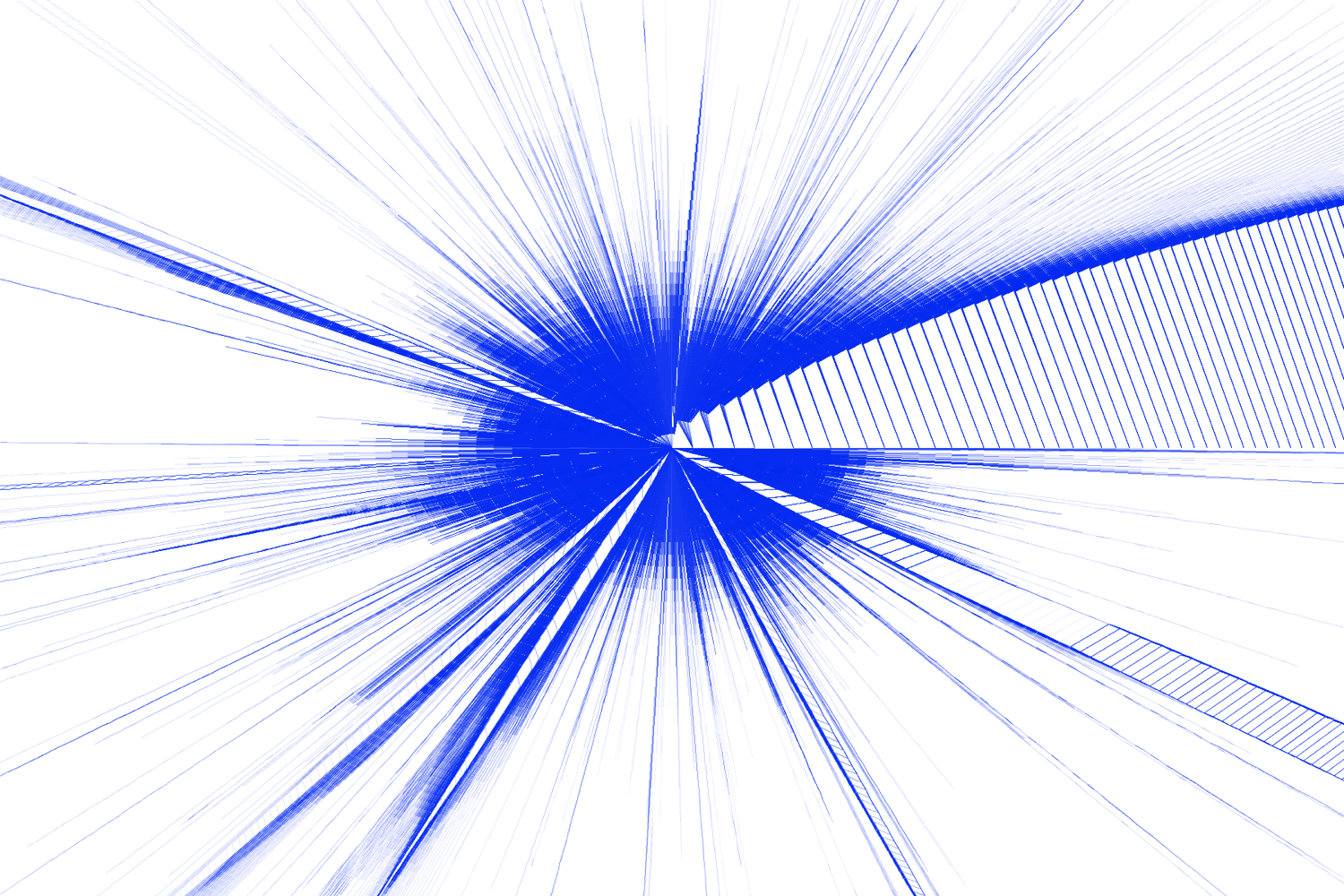}}
        }\\ %  ------- End of the first row ----------------------%
    \caption{%
      Example: Suffix tree constructed for {\it Aliser} malware artifact family data (79 files, 6,643,712 bytes). Trees are lexicographically ordered starting from the branch cut $0$ and winding counter clockwise to $2\pi$.  Notice that the bloom of wide branching and deep paths near argument 0 in the tree corresponds with substrings prefixed with null byte sequences; these are also low entropy strings.
     }%
   \label{fig:large_scale_data}
\end{figure}
\end{center}

\subsection{Traversal of Ukkonen's Suffix Tree to Create Clone Quantity Arrays.}
%For the traversal algorithms presented we utilized an externalized suffix tree augmented for corpus indexing based on Ukkonen's suffix tree.
Throughout the remainder of this section we assume a fixed corpus $\Omega$ with concatenated length $||\Omega||$ and number of artifacts $|\Omega|$, letting $\omega = \omega_0 \circ \hdots \circ \omega_{| \Omega |-1}$ be the concatenation of artifacts in the corpus.
The algorithm can generally be applied to any content map capable of emulating a suffix tree with the following minimum data fields for each node $\eta$ of the tree:  $\eta.O$ to access the offset in $\omega$, $\eta.C$ to access children of $\eta$, $\eta.L$ to measure the length of the branch in the suffix tree between $\eta$'s parent and $\eta$ (i.e. the length of string $\omega[\eta.O:(\eta.O+\eta.L)]$)\footnote{String $\omega[\eta.O:(\eta.O+\eta.L)]$ cooresponds with the state transition labels from $\eta$'s parent to $\eta$.}.  It is not necessary but beneficial to have a {\it suffix-link} $\eta.s$ pointing to the node representing the suffix of $\eta$, and for leaf nodes the {\it file identifier} $\eta.F$ and {\it local file offset} $\eta.o$.   

Let $\eta.C$ be the children of state $\eta$, sorted in order by the transition character (i.e. the order of $\Sigma$).  
We denote the $k$'th child (zero based index) of $\eta$ as $\eta.C[k]$, and assume the function: $\text{\sc{Child}}(\eta,k)$ which returns $\eta.C[k]$ if $k < |\eta.C|$ or $\emptyset$ otherwise. 
Note that $| \eta.C |$ is bounded by $|\Sigma|$ for all nodes of the suffix-tree. 
Let $root$ be the unique node not found as a child state for any other node.

Recall the correspondence of ${\cal L}(\Omega)$ and paths in the suffix tree: for suffix tree node $\eta$ we indicate this relation with $\bar \eta \in {\cal L}(\Omega)$ where $\bar \eta$ is the {\it path string} obtained by concatenating strings upon all branches from {\it root} to $\eta$. 

Throughout the traversal we maintain a stack\footnote{Should the stack grow to sizes beyond system memory, externalized data structures to support a large stack are possible.} of tuples denoted as ${\cal S}$.  
%Should the stack grow to sizes beyond system memory, externalized data structures to support a large stack are possible.
%We denote the stack as ${\cal S}$.
The tuples in the stack are of the following form:
$$
\langle \eta, k, l, T, z, \theta, \delta \rangle.
$$
With $\eta$ a unique node identifier, $k$ is a number between $0$ and $| \eta.C |$ indicating how many children of $\eta$ have been explored in post order, while $l$ represents the length of $\bar \eta$ and supports the computation of the {\it clone length} function $D( \bar \eta )$.
The variable $T$ represents the subset of corpus indices (file index) $\{ 0, 1, \hdots, n-1 \}$ indicating the covering files for $\bar \eta$ and supports the computation of the {\it clone file coverage} function $F( \bar \eta )$.  
The quantity $z$ counts the {\it clone multiplicity} function $C(\bar \eta)$ by computing the total number of leaf descendents of $\eta$ in the suffix tree.  The value $\theta$ is a vector over alphabet symbols in correspondence with $\Sigma$ supporting the computation of {\it clone entropy} function $H( \bar \eta )$.  
Let $\langle 0 \rangle_{\Sigma}$ be a count vector (over symbols of $\Sigma$) with all values initialized to $0$.  Finally $\delta$ charts the topological depth in the suffix tree counting the number of nodes between the root and $\eta$.

While the tree-traversal is a straightforward walk of the data structure, the ordering of computations needed to compute clone quantities $D, H, F, C$ for qualifying in set $\langle d, h, f, c \rangle$ must be sequenced carefully so we distributed them into {\sc PreOrderVisit} and {\sc PostOrderVisit} operations. 
We present the outline for traversal:
\vspace{-1em}
\begin{center}\fbox{
\parbox{0.98\linewidth}{%
\algo{
{\sc Traverse-Tree} \\
%\ \ \ 0:\ \ {\sc Push}(S, $root$), {\sc Push}(C ,{\sc Children}$(root)$), {\sc Push}(X, 0), {\sc Push}(A, $\O$), {\sc Push}(W, 0), {\sc Push}(H, $\langle 0 \rangle_{\Sigma}$ )\\
\ \ \ 0:\ \ {\sc Push}(${\cal S}, \langle root, 0, 0, \emptyset, 0, \langle 0 \rangle_{\Sigma}, 0 \rangle$ )\\
\ \ \ 1:\ \ $\phi \leftarrow \epsilon$  \\
\ \ \ 2:\ \ {\bf while} {\sc Length}(${\cal S}$) \\
\ \ \ 3:\ \ \ \ \ {\bf do} $\eta, k, l, T, z, \theta, \delta \leftarrow$ {\sc Pop}(${\cal S}$)\\
\ \ \ 4:\ \ \ \ \ \ \ \ \ $\mu \leftarrow ${\sc Child}$( \eta, k )$ \\
\ \ \ 5:\ \ \ \ \ \ \ \ \ {\bf if} $\mu \not = \emptyset $ \\
\ \ \ 6:\ \ \ \ \ \ \ \ \ \ \ \ {\sc Push}(${\cal S}$, $\langle \eta, k+1, l , T, z, \theta, \delta \rangle$ ) \\
\ \ \ 7:\ \ \ \ \ \ \ \ \ \ \ \ {\sc Push}(${\cal S}$, $\langle \mu, 0, l, \{\}, 0, \langle 0 \rangle_{\Sigma}, \delta + 1 \rangle$ )\\
\ \ \ 8:\ \ \ \ \ \ \ \ \ \ \ \ {\sc PreOrderVisit}( ${\cal S}, \phi$) \\
\ \ \ 9:\ \ \ \ \ \ \ \ \ {\bf else } {\sc PostOrderVisit}( ${\cal S}, \eta, l, T, z, \theta, \delta, \phi$) 
}
}%
}\end{center}
%Line 8 of {\sc Traverse-Tree} can be used to record a pre-ordering for the nodes of the tree; equivalently line 8 marks the first entry of node $n$ into the STACK completed in line $7$.  
While a node $\eta$ has additional children to explore it will be pushed back onto the stack with its child index incremented (line 6), and the $k$th child $\mu$ will be pushed immediately after (line 7).  When node $\eta$ has exhausted the exploration of children, flow-control reaches line 9 where quantities for the sub-tree rooted at $\eta$ will be aggregated upward to $\gamma$ (the parent of $\eta$).  In addition line 9 records a post-ordering of nodes in the tree, after which $\eta$ will not re-enter the stack again.  In addition {\sc PostOrderVisit} traverses the content of ${\cal L}(\Omega)$ in lexicographical order.
This outline completes the description of the traversal framework to compute clone quantities using suffix data structures.

Next we consider the \text{\sc{PreOrderVisit}} which provides the opportunity to initialize data for $\mu$ (child of $\eta$), extend the string $\phi$ with a contribution from $\mu$ to arrive at $\bar \mu$, and update $\theta$ needed to compute the entropy statistics $H(\bar \mu)$:
\vspace{-1em}
\begin{center}
\fbox{
\parbox{0.98\linewidth}{%
\algo{
{\sc PreOrderVisit}(${\cal S}, \phi$) \\
\ \ \ 0:\ \ $\mu, k, l, T, z, \theta, \delta $ $\leftarrow$ {\sc Top}(${\cal S}$)\\ 
\ \ \ 1:\ \ {\bf if} ( \sc{Leaf}($\mu$) ) \\
\ \ \ 2:\ \ \ \ \ $z \leftarrow 1$ \\
\ \ \ 3:\ \ \ \ \ $T \leftarrow T \cup \ \{ \mu.F \}$ \\%{\sc Push}( A, {\sc Pop}( A ) $\cup$ $\{$ s[I] $\}$ )
\ \ \ 4:\ \ \ \ \ $l \leftarrow -1$ \\
\ \ \ 5:\ \ {\bf else} \\
\ \ \ 6:\ \ \ \ \ $l \leftarrow l + ( \mu.L )$ \\
\ \ \ 7:\ \ \ \ \ $\lambda \leftarrow \omega[ \mu.O: ( \mu.O + \mu.L) ]$ \\
\ \ \ 8:\ \ \ \ \ $\theta \leftarrow \theta  + \langle \lambda \rangle_{\Sigma}$ \\
\ \ \ 9:\ \ \ \ \ $\phi \leftarrow \phi \circ \lambda$ %\\ \% record the string $\tilde s$ by concatenation of $\bar s$ onto $\phi$.\\
%\ \ \ 10:\ \ \ \ \ {\sc Push}( ${\cal S} , \langle \mu, k, d, T, c, \theta \rangle$ ).
}
}%
}\end{center}
In line 0 of {\sc PreOrderVisit} we access $\mu$'s variables stored at the stack's top (line 7 of {\sc Traverse-Tree}).  The function {\sc Leaf} may be implemented by checking the predicate:  $(|\mu.C|=0)$.
In lines 2-4 we treat the case when $\mu$ is a leaf: quantities {\it clone multiplicity} $z$ and {\it clone file cover} $T$ are initialized and later will be aggregated upward during the {\sc PostOrderVisit}, and quantity $l$ is set to $-1$ to indicate a suffix of $\omega$ and could be interpreted as ``read to the end of corpus.''  
Lines 6-9 handle computations required for internal branch nodes; these include updating the string $\phi$ from $\bar \eta$ to $\bar \mu$ (line 9) by extracting the string from the corpus associated with the suffix link between $\mu$ and $\mu$'s parent $\eta$ (Line 7).  
Updating the depth variable $l$ from $| \bar \eta |$ to $| \bar \mu |$ is performed in line 6, and updating a running symbol count of $\phi$ is performed in line 8.  

Next we consider the \text{\sc{PostOrderVisit}} providing the last opportunity to perform computations obtained from node $\eta$:
\vspace{-1em}
\begin{center}\fbox{
\parbox{0.98\linewidth}{%
\algo{
{\sc PostOrderVisit}(${\cal S}, \eta, l, T, z, \theta, \delta, \phi$) \\
\ \ \ 0:\ \ $\gamma, k_\gamma, l_\gamma, T_\gamma, z_\gamma, \theta_{\gamma}, \delta_{\gamma}$ $\leftarrow$ {\sc Top}(${\cal S}$) \% parent of $\eta$ is $\gamma$ \\ 
\ \ \ 1:\ \ $z_{\gamma} \leftarrow z_{\gamma} + z $ \\
\ \ \ 2:\ \ $T_{\gamma} \leftarrow T_{\gamma} \cup T$ \\
\ \ \ 3:\ \ \sc{Print}( $\eta, \eta.O, \eta.L, \eta.F, \eta.s, \delta, \langle l, h( \frac{1}{z}\theta ), T, z \rangle$ ) \\
\ \ \ 4:\ \ $\phi \leftarrow \phi[0:l_{\gamma}]$ \% return string to $\bar \gamma$. % by truncate string $\phi$
}
}%
}\end{center}
Notice that in line 3, the printing of $\langle l, h( \frac{1}{z}\theta ), T, z \rangle$  are evaluations of $D( \bar \eta )$, $H( \bar \eta )$, $F( \bar \eta )$, $C( \bar \eta )$.
Line 2 and 3 of \text{\sc{PostOrderVisit}} aggregate $z$ the {\it clone multiplicity} and compute the {\it clone file cover} set $T$ currently held by $\eta$ upward to $\gamma$ ($\eta$'s parent) at the top of the stack at the time {\sc PostOrderVisit} is called (note that only a parent $\gamma$ can precede a child $\eta$ in stack insertion: see lines 6-7 of {\sc Traverse-Tree}).  
Line 3 writes the enhanced array to output and could provide an opportunity to conduct further and more general analysis for clone membership.  
Finally line 4 reduces the current content string $\phi$ to $\bar \gamma$ by truncating $\eta.L$ symbols from $\bar \eta$.

\begin{lem}
{\sc Traverse-Tree} is $O( || \Omega || \log{( |\Omega | )} )$.
%in the size of {\sc{STACK}} and $|\omega|$ (i.e. constant in the length of stack size and corpus length)
\end{lem}
{\it Proof}: The total number of $\text{\sc{Pop}}$'s (line 3 of {\sc Traverse-Tree}) is bounded by twice the number of edges in the suffix tree and therefore bounded by $4||\Omega||$ as the maximum number of edges in a tree which is less than twice the number of leaf nodes.  Therefore, the loop is performed $O(|\omega|)$ times and the complexity consideration is reduced to that of $\text{\sc{PreOrderVisit}}$ and $\text{\sc{PostOrderVisit}}$.  The traversal guarantees that {\sc PreOrderVisit} and {\sc PostOrderVisit} are called once per node.

Line 3 of \text{\sc PreOrderVisit} and lines 2 and 3 of \text{\sc PostOrderVisit} are $\log{(||\Omega||)}$ set operations.  
Lines 7-9 of {\sc PreOrderVisit} can be analyzed by amortizing  across all nodes of the tree during traversal, since the load size of $\lambda$ over all nodes of the tree is no greater than loading the entire corpus.  The total cost of all operations is therefore bounded by $O(||\Omega||)$.   
All other operations in \text{\sc{PreOrderVisit}} and \text{\sc{PostOrderVisit}} are a $O(1)$.  Therefore the entire runtime is bounded by $O(||\Omega|| \log{(|\Omega|)} )$.  
$\clubsuit$

{\bf Note:}  More generally line 3 of {\sc PostOrderVisit} could be replaced by any method that is $O(1)$ in the depth of stack ${\cal S}$ and $O( \log{ (|\Omega|) })$ to get a slightly stronger lemma allowing for additional analysis involving the relation between a node $\eta$ and its parent $\gamma$, or some limited size ancestral chain, for example: $\eta, \eta.P, \eta.P.P$.

\subsubsection{Enhancing a Suffix Array with Clone Quantities.}
With the runtime for {\sc Traverse-Tree} resolved as $O( || \Omega || \log{( | \Omega | ) } )$ we now focus on the transformation of data that line 3 of {\sc Post-Order-Traversal} yields.  While it produces a tabulated form of data that allows us to test node membership in $\langle d,h,f,c \rangle$, it also achieves a full map of all suffixes printed in lexicographical order thereby creating an {\it enhanced suffix array} augmented with quantities of {\it clone length} $D$, {\it clone entropy} of $H$, {\it clone file coverage} $F$ and {\it clone multiplicity} $C$.

In the subsequent section we show how this map can be used to support measures leading to pairwise distance based on clones in common and clustering based on common clones.  
We suggest the outline above as a framework to extend notions of clones further; for example, measuring the distance to specific symbol frequency vectors such as a topic vector or witness complex \cite{DeSilva:2004}, which we plan to pursue as future work.

\subsection{Max-Clone Representation.}
Clone sets have an inherent redundancy which displays perplexing patterns related to the self similarity of the suffix tree data structure.  To simplify matters we describe a representation of a set $\langle d,h,f,c \rangle$ that is both easy to visualize and minimal in that it selects the smallest subset of {\it representatives} from $\langle d,h,f,c \rangle$ for which all other members are suffixes of a representative with equal value for $F$ and $C$. 
We shall argue that knowing all members provides no additional information beyond knowing the representation.  We provide an indication of the type of reductions the representation offers in practice.
%Next we provide a definition of a suffix-relation, provide two methods for enumerating the max-clone representation of the clone class as algorithms.  
%
%% We defer the considerations of minor issues arising from representing suffix-data by suffix trees to a longer paper, but must indicate that our method is a clone sampler with the possibility of missing the identity of a true clone (false-negative) due to entropy variations on long branches (in the implicit states).  In practice it appears to be a low-probability event and a minor concern in the applications we present.
%For each node of the externalized compressed suffix tree $\mu$ the string $\bar \mu$ represents the string from {\it root} to $\mu$ can be tested for inclusion in a particular clone class $\langle d,h,f,c \rangle$ either as part of a modified version {\sc Statistical-Analysis-Identify} based on {\sc statistical-Analysis} or by analyzing the output of {\sc Statistical-Analysis} as developed above, either option represents a constant time operation applied to each node or line of text.  
%Therefore its relatively straight forward to identify the nodes of a suffix tree that encode clones of a given class, and can be performed in linear time $O(|\omega\|)$.  

Two nodes in the externalized suffix tree are suffix-related\footnote{Ukkonen's construction includes suffix links.}, denoted $\mu \prec \rho$, if $\bar \rho = x \bar \mu$ for some symbol $x \in \Sigma$.
Given a specific clone set ${\cal B} = \langle d,h,f,c \rangle$ we can extend the suffix-relation $\prec$ to members with a {\it level-set-suffix-relation} $\prec_{\cal B}$ on all nodes of the tree as: 
$$
( \mu \prec_{\cal B} \rho ) \Leftrightarrow (\mu \prec \rho ) \wedge ( \bar \mu \in {\cal B} ) \wedge ( \bar \rho \in {\cal B} ) \wedge ( F(\bar \mu) = F(\bar \rho) ) \wedge (C( \bar \mu ) = C( \bar \rho )).  
$$
We define the {\bf max-clone representation} of a clone class $\langle d,h,f,c \rangle$ as the strings associated with maximal elements of the relation $\prec_{\langle d,h,f,c \rangle}$, and we denote the {\it max-clone representation} as $\langle \langle d,h,f,c \rangle \rangle$.
%We present two methods for computing this representation both linear time in $|\omega|$.  
%The first method we call {\sc Tiling} as it uses the output {\sc Statistical Analysis} and computes a maximum covering set of clones on a model of the data, the second method utilizes the {\it suffix links} from Ukkonen's algorithm and allows for the {\it maximum-clone representation} to be computed on the compressed suffix tree by slightly modifying the post ordering to prioritize the traversal using suffix links.
Computing the {\it max-clone representation} is easily seen to be $O(|| \Omega || )$; see the appendix for a graph algorithm that computes the max-clone representation.

\paragraph{Justification:}  The suffix relation is a particularly appropriate order for reducing the representation (to maximal clones), because any non-representative member of the clone class is a suffix of a representative member with identical values of $F,C$ (level set).    Furthermore in applications we argue that this representation translates directly to the longest common strings of interest in data and we provide examples of how the max-clone representation may be visualized in figure \ref{fig:visulaization_duqu_stux} for {\it Duqu} and {\it Stuxnet} malware data.

{\bf Reduction in practice:}  The max-clone representation is an effective data reduction in practice.
%These reductions are level sets in selected parameters of cloneage (we let $h$ vary freely so long as it stays above a given threshold).  
%The more parameters of the clone model, the more we may expect the the representation to grow as well.   
In figure \ref{fig:visulaization_duqu_stux} we present a visualization of a max-clones for $\langle 1000, 2.0, 2, 2 \rangle$.  
In this case the number of nodes of the suffix tree quantified by $\langle 1000, 2.0, 2, 2 \rangle$ is 25,177, yet the max-clone representation comprises $7$ clones located at 17 offsets in the corpus.  
The max-clone representation signals what and where largest relations in data may be found.  
In this case the max-clone representation selects $2.780868 \times 10^{-4}$ fractional amount of clones from $\langle 1000, 2.0, 2, 2 \rangle$. % and in our opinion captures the major features of shared data which an investigation of provenance can utilize.
\paragraph{Conclusion:}  For corpus $\Omega$ the max-clone representation for $\langle d,h,f,c \rangle$ denoted $\langle\langle d,h,f,c \rangle\rangle$ is computable in $O( ||\Omega|| \log{ ( | \Omega | )})$ by first building suffix-data structures, traversing the suffix-data with {\sc Traverse-Tree}, and identifying maximal elements of $\prec_{\langle d,h,f,c \rangle}$.  

\section{Applications and Results.}
The remainder of the paper focuses on applications of our methodology to malware artifact data.  We address the motivating problems and illustrate results on actual malware data artifacts.  
We focus on the problem of {\it Stuxnet} and {\it Duqu} (which represents a difficult challenge in cyber security) and show the use of clone sets to identify and measure the evidence for provenance.
Using the max-clone representation $\langle \langle d,h,f,c \rangle \rangle$, we sketch how to construct Jaccard similarity coefficients to compare artifacts in a pairwise manner.  
We present Jaccard coefficients for this problem and the results indicate that the relation between the {\it Stuxnet} and {\it Duqu} malware sets signals an overlap detectable with the Jaccard coefficient.
Finally we consider the Jaccard similarity coefficients for cyber secruity data and provide experimental designs in terms of coverage and compression.

\subsection{Mapping Clone Features and Visualization. }
\begin{figure}[h!]
  \centering
    {%
      \includegraphics[angle=90,width=88mm]{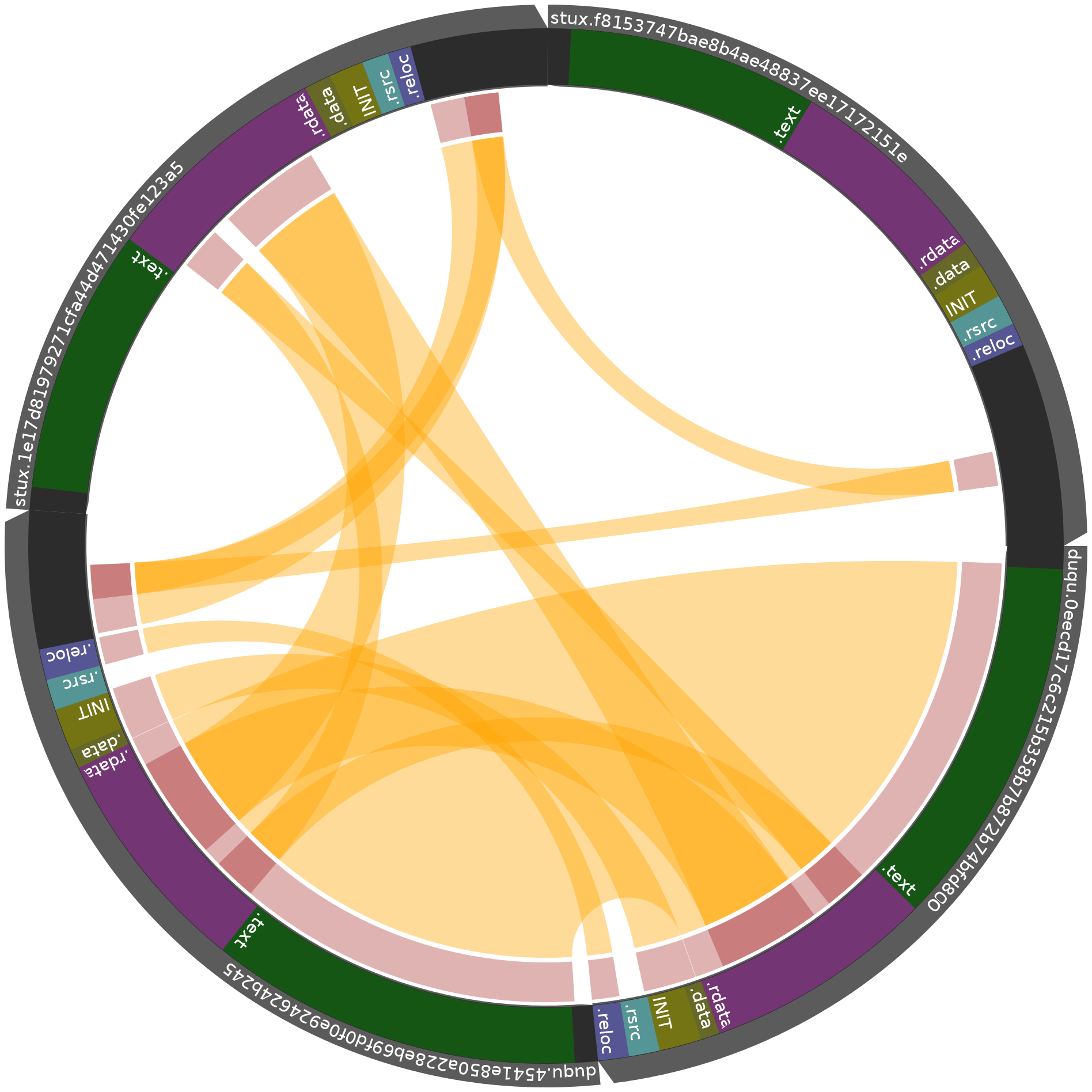}%clone_map__2_0_1000_2.png}
    }
  \caption{\small Example: {\it Duqu} vs {\it Stuxnet}: visualization of $\langle \langle 1000, 2.0, 2, 2\rangle \rangle$. Binaries of four malicious code samples (two from the {\it Duqu} family and two from the {\it Stuxnet} family) are illustrated as regions of an annulus.  Small notches in the outer circumference mark the beginning of a binary and can be viewed at approximate angles of: $\frac{\pi}{2}, 0$ for the {\it Duqu} samples and $\frac{3\pi}{2}, \pi$ for the {\it Stuxnet} samples.  Next files are divided into {\it sections} and color coded by section name {\it .text, .rsrc, .rdata, .data, INIT, reloc}.   
Max-Clones from $\langle 1000, 2.0, 2, 2 \rangle$ are illustrated as counter-arcs passing through the interior region and connecting orthogonally to the annular region representing the binary layouts.  These counter-arcs show the locations and length of max-clones when $c>1$.  An annular region, just interior to the view of file formats, illustrates the copy number of each cloned region with a red alpha channel.  Partial transparency (alpha channel) helps with signaling clone matches contained as substrings to larger matching clones.
}
\label{fig:visulaization_duqu_stux}
\end{figure}

In Figure \ref{fig:visulaization_duqu_stux} we assemble a set of binary artifacts (the driver artifacts) matching anti-virus signatures for either {\it Duqu} or {\it Stuxnet} malware groups as studied in \cite{Chien:2012},\cite{stuxnet:2010} where reverse engineering techniques discovered evidence supporting the hypothesis fo a common {\it provenance} or history of development.  
While these discoveries were important to the security community and also (from 2012 forward) to the mainstream media, the question of identifying all the evidence supporting the findings remained open.  
%Under the assumption that code-clones arise with common provenance, we illustrate methods that identify and map the evidence (of the underlying {\it code-clones}) supporting the claim that two binaries derive from common source.
%\begin{table} 
%\begin{center}
%\begin{tabular}{|l|l|l|l|}
%\hline
%binary artifact & id & size & sections \\
%\hline
%\hline
%duqu.0e & 1 & 24960 & 6  \\%{\small .text, .rdata, .data, INIT, .rsrc, .reloc}\\
%duqu.45 & 2 & 29568 & $6^*$ \\%{\small .text, .rdata, .data, INIT, .rsrc, .reloc slack }\\
%stux.1e & 3 & 25552 & $6^*$ \\%{\small .text, .rdata, .data, INIT, .rsrc, .reloc slack}\\
%stux.f8 & 4 & 26616 & $6^*$ \\%{\small .text, .rdata, .data, INIT, .rsrc, .reloc slack}\\
%\hline  
%\end{tabular}
%\caption{The {\it Duqu}-{\it Stuxnet} data set. All binaries identify six sections {\it .text, .rdata, .data, INIT, .rsrc, .reloc} in the PE header; $6^*$ indicates the existence of data beyond the last %section. Data constructs a suffix tree with $174,208$ tree nodes of which $67,517$ are internal nodes (the maximum number of clones with $c\geq2$).}
%\label{tab:datasetduquvsstux}
%\end{center}
%\end{table}
%%%%%%%%%%%%%%%%%%%%%%%%%%%%%%
%\vskip{ -420mm }
\paragraph{Conclusions:} The visualization of data is useful for data triage and establishing priorities for costly reverse engineering resources.  
For example in the image \ref{fig:visulaization_duqu_stux} a high entropy string of sizable length in found in the slack section\footnote{Slack sections are areas of data not reported by the program's load table.} of binaries including both {\it Stuxnet} and {\it Duqu}.

\subsection{Measurements of Shared Clone Features. }

Using Figure \ref{fig:visulaization_duqu_stux}, which illustrate $\langle \langle 1000, 2.0, 2, 2 \rangle \rangle$, we can indentify and count the distinct number of clones as $7$ max-clones with $14$ distinct offsets in the corpus.  
In this section we develop the Jaccard similarity coefficient to measure the percentage of content in common (given a clone class) in pairs of files.  We further illustrate how these measures may vary on the clone class $\langle d,h,f,c \rangle$.

Fixing the Corpus $\Omega = \{\omega_0, \hdots, \omega_{n-1} \}$ and given a clone set $\langle d, h, f, c \rangle$, A Jaccard similarity coefficient for all pairs of artifacts is fairly straightforward and is computed as follows:  
For any subset $I \subset \{ 0, \hdots, n-1 \}$, identify all the clones which have a region contained in all of the artifacts of $I$, so assume the function:
$$
\text{{\sc Cover }}( I ) = \{ \lambda \in \langle \langle d, h, f, c \rangle \rangle : I \subset \text{{\sc File}}( \lambda^{-1} ) \}
$$
with $\text{{\sc File}}( \lambda^{-1} ) = \{ \text{{\sc File}}( r ) : r \in \lambda^{-1} \}$.  
For comparison of artifact $i$ against a subset $I$ we count the number of indices of $\omega_i$ covered by strings from $\text{{\sc Cover }}( I )$.
$$
A( i, I ) = \sum_{a=0}^{|\omega_i|} \phi ( \omega_i[a:], \text{{\sc Cover}}( I ) )
$$
with: 
$$
\phi( \omega[a:], S ) = \begin{cases} 1 \text{ if } \exists b: \omega[a:b] \in S \\ 0 \text { o.w. } \end{cases}
$$
We introduce the Jaccard similarity coefficient as $J( I )  = \frac{ \sum_{i \in I} A( i, I ) }{ \sum_{\i \in I} |\omega_i| }$ and interpret this as the percentage of a subset covered by the given clone set $\langle d,h,f,c \rangle$.  
{\bf Pairwise Measures:} Table \ref{tab:jaccard} we compute the {\sc Pairwise-Jaccard} by considering subsets $I$ with $|I| = 2$.  The pairwise measures are presented for a range of different clone sets to give a sense of measure dependencies on clone quantities $d,h$.  

\begin{figure}[!ht]
\centering{
  \begin{tabular}{|r|l|l|l|l|l|l|}
  \hline
  \multicolumn{7}{|c|}{Jaccard similarity coefficient} \\
  \hline
  \multicolumn{1}{|c|}{clone-class $\langle d,h \rangle$} & \multicolumn{6}{c|}{} \\
  \cline{1-1}
  \multicolumn{1}{|c|}{ {\small $\langle 10, 0.25\rangle$  $\langle 1000, 0.25\rangle$} } & \multicolumn{6}{c|}{binary} \\
  \cline{2-7}
  {\small $\langle 10, 2.0 \rangle $ $\langle 1000, 2.0 \rangle$ } & \multicolumn{2}{c|}{duqu.45} & \multicolumn{2}{c|}{sutx.1e} & \multicolumn{2}{c|}{stux.f8} \\
  \hline 
  \hline
    \multirow{2}{*}{ duqu.0e} & 0.91 & 0.86 & 0.41 & 0.22 & 0.25 & 0.00 \\
  \cline{2-7}
                            & 0.87 & 0.86 & 0.31 & 0.22 & 0.14 & 0.00 \\
  \cline{1-1} \cline{2-7}
    \multirow{2}{*}{ duqu.45} &  \multicolumn{2}{c|}{} & 0.51 & 0.29 & 0.36 & 0.05 \\
  \cline{4-7}
                             &  \multicolumn{2}{c|}{} & 0.42 & 0.29 & 0.25 & 0.05 \\
  \cline{1-1} \cline{4-7}
   \multirow{2}{*}{ stux.1e} & \multicolumn{2}{c}{} & \multicolumn{2}{c|}{} & 0.57 & 0.05 \\
  \cline{6-7}
                            & \multicolumn{2}{c}{} & \multicolumn{2}{c|}{} & 0.47 & 0.05 \\
  \hline
  \end{tabular}
  } %\hfill 
 \caption{Jaccard coefficients for pairwise binaries in {\it Duqu}-{\it Stuxnet} data set for clone classes $\langle d,h,f,c \rangle$, with $f=2,c=2$. (b) Fractional amount of all data covered by a clone from $\langle d,e ,2,2 \rangle$ for various values of $d,e$.}
 \label{tab:jaccard}
\end{figure}

\begin{figure}[!ht]
  \centering 
  \includegraphics[angle=0,width=88mm]{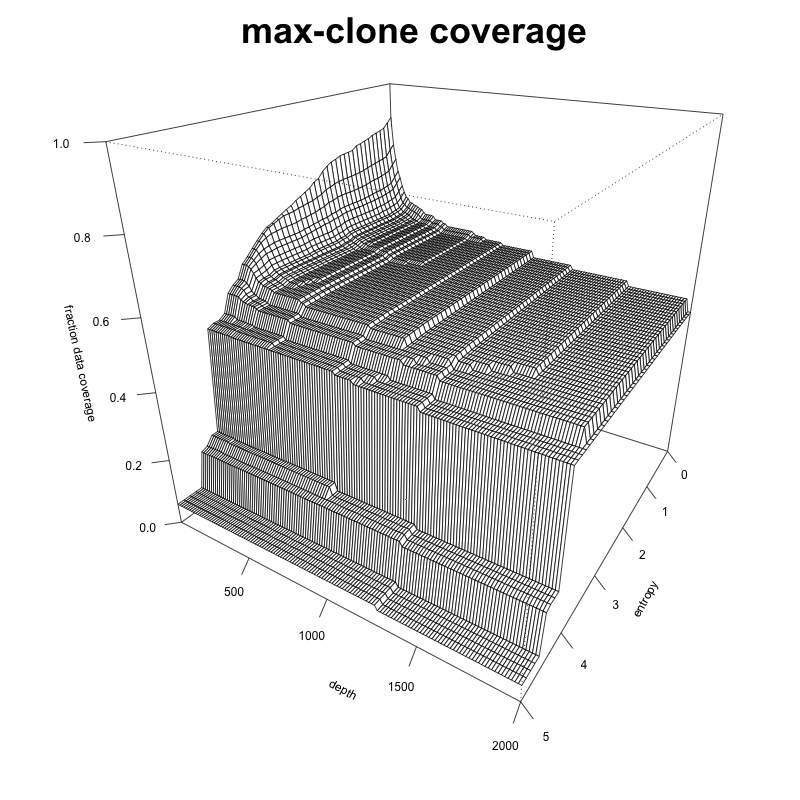} 
  \caption{Coverage:  fractional amount of all data covered by a clone from $\langle d,e ,2,2 \rangle$ for various values of $d,e$.}
  \label{fig:coverage}
\end{figure}

\paragraph{Conclusions:}  The Jaccard index applied to pairs of files such as in the experiment with {\it Stuxent} and {\it Duqu} binary files may indicate shared provenance or present evidence that shared provenance is a candidate mode for binaries with unknown histories.  In the above computation using clone class $\langle 1000,2,2,2 \rangle$ the measure of 29\% pairwise identity across the family boundary turns out to be a significant amount of clone features.  Further the measure can be applied to incoming samples and measured against a known dictionary of examples.  Results from \ref{fig:coverage} provides useful information on how to set parameters for dictionary matching.

{\bf Set Algebra Measure:} 
We construct a mixed data set including binaries from four malware families including: {\it Duqu, Poison Ivy, Stuxnet}, and {\it Zeus/Zbot}, and binaries from two operating systems:  {\it Linux} and {\it Win7}.  In table \ref{tab:datatriage} we list the mixed data set used in this experiment.  In the experiment we let $I = \{1, \hdots, 16\}$ be the artifact index with artifact cluster identity withheld. Using the clone set $\langle 80, 0.6, 2, 2\rangle$ we enumerate subsets of $A \subset I$ with non-empty $\text{\sc Cover}( A )$.   For each subset of $A \subset I$ we may compute the Jaccard similarity coefficient $J( A )$, and in Table \ref{tab:data-family-okay} we present the rank ordering of the result.

\begin{figure}[!ht]
\centering
{\scriptsize
 \subfigure[Data]{
     \label{tab:datatriage}
      {\begin{tabular}{|l|l|l|}
\hline
binary artifact & id &  artifact cluster \\
\hline
\hline
duqu.0e & 1&  \multirow{2}{*}{\it Duqu} \\
duqu.45 & 2&   \\
\hline
linux.bzip2 & 3 &  \multirow{5}{*}{\it Linux} \\
linux.pwd & 4 &  \\
linux.sed  & 5 &  \\
linux.su & 6 &   \\
linux.tar & 7 &   \\
\hline
pi.0a..67 & 8 &  \multirow{2}{*}{\it Poison Ivy}  \\
pi.0a..cf & 9 &  \\
\hline
stux.1e..a5 & 10 &  \multirow{2}{*}{\it Stuxnet} \\
stux.f8..1e & 11 &   \\
\hline
win7.calc & 12 &  \multirow{3}{*}{\it Win7}  \\
win7.shutdown & 13 &  \\
win7.soundrecorder & 14 &   \\
\hline
zbot.20..f6 & 15 &  \multirow{2}{*}{\it Zeus} \\
zbot.a8..8e & 16 & \\
\hline  
\end{tabular}}
    }
    \hfill
    \subfigure[Discovered Topics]{%
     \label{tab:data-family-okay}
      {\begin{tabular}{|l|l|l|l|}
\hline
   & subset & number of  & comment \\
J(A) & $A \subset I$ & clones &  \\
\hline
\hline
0.882959 &	            \{1,2\} & 	7 & {\it Duqu} \\
\hline
0.819336 &	            \{8,9\} & 	10 & {\it Poison Ivy} \\
\hline
0.268531 &	         \{1,2,10\} & 	9 & \multirow{3}{*}{\it Duqu vs Stuxnet}\\
0.122605 &	           \{2,10\} & 	6 & \\ 
0.077236 &	        \{2,10,11\} & 	7 & \\
\hline
0.076004 &	          \{10,11\} & 	17 & {\it Stuxnet}\\
\hline
0.036384 &	      \{1,2,10,11\} & 	9 & {\it Duqu vs Stuxnet}\\
\hline
0.028313 &	            \{4,7\} & 	22 & \multirow{6}{*}{\it Linux}\\
0.015117 &	            \{3,4\} & 	2 & \\
0.013848 &	            \{3,5\} & 	4 & \\ 
0.013570 &	            \{3,6\} & 	2 & \\
0.009218 &	            \{5,6\} & 	1 & \\
0.007921 &	      \{3,4,5,6,7\} & 	4 & \\
\hline
0.007230 & 	          \{13,14\} & 	5 & {\it Win7} \\
\hline
0.004880 &	            \{4,6\} & 	1 & {\it Linux}\\
\hline
0.004104 &	         \{1,2,11\} & 	1 & {\it Duqu vs stux}\\
\hline
0.003735 &	          \{4,6,7\} & 	2 & \multirow{3}{*}{\it Linux}\\
0.003445 & 	        \{3,5,6,7\} & 	2 & \\
0.003074 &	            \{6,7\} & 	1 & \\
\hline
0.002600 &	       \{12,13,14\} & 	7 & \multirow{2}{*}{\it Win7}\\
0.002521 &	          \{12,13\} &	6 & \\
\hline  
\end{tabular}}
    }
    \caption{Experiment using random artifacts shows that the measure $J(A)$ for $A \subset I$ presents good recovery options for artifact triage.  (a) Data set: a mixture of random samples from several distinct sets. (b) Rank by J(A), (top 21 entries) with measure $\geq 0.025$ shown. }
    \label{fig:setalgres}
}
\end{figure}
\paragraph{Conclusions:}
The results reported in the experiment above indicate the usefulness of applying these measures to unknown data for triage or a first order pass to identify topics in data sets.  
While we defer a statistical treatment to a future effort the result above is useful in indicating the significance of the \it{Duqu}, \it{Stuxnet} comparison and also indicates the expectation of increased measures of clones in common in binaries chosen randomly from related activities.  

\subsubsection*{Acknowledgments.} 
We would like to thank the Members of Software Engineering Institute:  Chuck Hines, Jeffrey Havrilla, Leigh Metcalf and Rhiannon Weaver for the many discussions about cyber security science. 
The research reported here was supported by CMU SEI line funded research program.

\bibliographystyle{named}
\bibliography{ref_paper}

\end{document}